\documentclass[pre,twocolumn,amsmath,amssymb,floatfix,nofootinbib]{revtex4}
\usepackage{mathrsfs,bm}
\usepackage{longtable,lscape}
\usepackage{txfonts}
\usepackage{indentfirst}
\usepackage{graphicx,color,dcolumn,booktabs}
\usepackage{multirow}
\usepackage{indentfirst}
\usepackage{multirow}
\usepackage{epsfig}
\usepackage{graphicx,color,dcolumn}
\usepackage{epstopdf}
\usepackage{amsmath}
\usepackage{multirow}
\usepackage{diagbox}
\usepackage{changes}
\usepackage{threeparttable}
\usepackage{enumitem,overpic}

\usepackage{color}
\usepackage{amssymb}
\definecolor{cover}{rgb}{0.77,0.87,0.88}
\definecolor{blueone}{rgb}{0.1,0.1,.7}
\definecolor{citec}{rgb}{0.14,0.47,0.09}
\definecolor{two}{rgb}{0.0,0.5,0.}
\definecolor{three}{rgb}{.5,.1,0.15}
\usepackage[bookmarks=true,bookmarksopen=false,plainpages=false,breaklinks=true,
   bookmarksnumbered=true,hypertexnames=false,
   filecolor=blue,urlcolor=three,menucolor=three,
   linkcolor=three,citecolor=blueone, colorlinks,
   anchorcolor=blue,runcolor=pink,frenchlinks=red
   pdfstartview=FitH,pdftitle=title,%
   pdfauthor=author]{hyperref}

\usepackage{relsize}
\usepackage{xspace}
\def\babar{\mbox{\slshape B\kern-0.1em{\smaller A}\kern-0.1em
    B\kern-0.1em{\smaller A\kern-0.2em R}}}

\allowdisplaybreaks[4]
\usepackage{color}
\begin{document}

\title{Calcium oscillation on homogeneous and heterogeneous networks of
ryanodine receptor}

\author{Zhong-Xue Gao\footnote{These authors have contributed equally to this
work.\label{aa}}, Tian-Tian Li\footnotemark[1], Han-Yu
Jiang\footnote{Corresponding author: jianghy@njnu.edu.cn} and Jun
He\footnote{junhe@njnu.edu.cn}}

\affiliation{School of Physics and Technology, Nanjing Normal University,
Nanjing 210097, China}

\date{\today}

\begin{abstract}

Calcium oscillation is an important calcium homeostasis, imbalance of which is
the key mechanism of initiation and progression of many major diseases.  The
formation and maintenance of calcium homeostasis are closely related to the
spatial distribution of calcium channels on endoplasmic reticulum, whose complex
structure was unveiled by recent observations with super resolution imaging
techniques.  In the current paper, a theoretical framework is established by
abstracting the spatial distribution  of the calcium channels as a nonlinear
biological complex network with calcium channels as nodes and Ca$^{2+}$ as
edges. A dynamical model for a ryanodine receptor (RyR) is adopted  to investigate
the effect of spatial distribution on calcium oscillation. The mean-field model
can be well reproduced from the complete graph and dense Erd\"os-R\'enyi
network. The synchronization of RyRs is found important to generate a global
calcium oscillation. Below a critical density of the Erd\"os-R\'enyi or
BaraB\'asi-Albert network, the amplitude and interspike interval decrease
rapidly with the end of disappearance of oscillation due to the desynchronization.
The clique graph with a cluster structure can not produce a global oscillation due
to the failure of synchronization between clusters.  A more realistic geometric
network  is constructed in a two-dimensional plane based on the experimental
information about the RyR arrangement of clusters and the frequency distribution
of cluster sizes. Different from the clique graph, the global oscillation can be
generated with reasonable parameters on the geometric network. The simulation
also suggests that existence of small clusters and rogue RyR's plays an important
role in the maintenance of  global calcium oscillation through keeping
synchronization between large clusters.  Such results support the heterogeneous
distribution of RyR's with different-size clusters, which is helpful to
understand recent observations with super resolution nanoscale imaging
techniques. The current theoretical framework can also be extent to investigate
other phenomena in calcium signal transduction.

\end{abstract}

\maketitle
\section{INTRODUCTION}\label{sec1}

Calcium homeostasis is the key mechanism to maintain life-sustaining activities. Its imbalance
is responsible for the  initiation of  many major diseases, such as
neurodegenerative and cardiovascular
diseases~\cite{Berridge2016,Terentyev2008,Fiedler2002}. As an important calcium
homeostasis, calcium oscillation is a ubiquitous signal in all cells, and provides
efficient means to transmit intracellular biological information through its
amplitude and frequency, therefore, attracting much
attention~\cite{Cohen2002,Wei2021,Sneyd2017,Smedler2014}.  Many models have
been proposed to study the calcium signaling transduction and the formation of
calcium
homeostasis~\cite{DeYong1992,Keizer1997,Sneyd2002,Cheng2008,Rudiger2010,Ruediger2012}.
In the existing models, the calcium channels, such as ryanodine receptors (RyR's) and inositol
1,4,5-trisphosphate receptors (IP$_3$R), play an important role in the regulation of
intracellular calcium concentration. After inclusion of other mechanisms, such
as  calcium bump, calcium leak, and mitochondrion, the phenomena of calcium
homeostasis, including calcium oscillation, were reproduced from many proposed
models to explain  experimental
observations~\cite{DeYong1992,Keizer1997,Fall2001,Maral2000,Szopa2013}. 

Most  existing models were constructed under the mean-field
ansatz~\cite{DeYong1992,Keizer1997,Sneyd2002,Cheng2008}. In such models, the
calcium concentrations near all calcium channels are set to be the same value
and all channels behave in the same manner. It requires homogeneous distribution
of the channels  on the endoplasmic reticulum, which obviously conflicts with
the experimental observations that the calcium channels exhibit a cluster
structure composed of several to tens of 
channels~\cite{Franzini-Armstrong1999,Baddeley2009,Smith2009,Jayasinghe2018}.
Such a heterogeneous cluster structure makes the intracellular calcium signaling
exhibit obvious hierarchy. Many models have been proposed to simulate such a
structure~\cite{Ruediger2012,Rudiger2010,Cao2013,Walker2014,Walker2015}. For
example, using deterministic-stochastic simulations, R\"udiger $et\ al.$ studied local calcium signaling in a cluster with compact
distribution of the channels in a rectangle, and well described calcium puffs in
neuronal cells~\cite{Rudiger2010}.

These studies about the cluster structure deepen our understandings of the
hierarchy of calcium signaling. However, recent super resolution observations
suggest that the clusters have a more complex internal structure.  In many studies,
the cluster is assumed to be packed
compactly~\cite{Rudiger2010,Mesa2021,Cao2013,Walker2014,Walker2015}, forming a
tight lattice as suggested by early experimental
observations~\cite{Franzini-Armstrong1999,Baddeley2009}, which means that the
mean-field ansatz is still suitable to study the local dynamics in a cluster, at
most with some small modifications. However, recent super-resolution nanoscale
imaging shows that the accurate characteristics of calcium-channel arrangement
in a cluster are not trivial and should be considered
seriously~\cite{Scott2021,Jayasinghe2018}. The cluster is not compact, has
irregular shapes and different sizes with randomness. In
Refs.~\cite{Iaparov2019,Iaparov2021}, activation of RyR's was found to be sensitive to
the arrangement of RyR's in a cluster.  Their study suggests that a regular
compact picture of cluster and mean-field approximation are not suitable for
simulation of RyR arrangement observed experimentally.  The RyR arrangement
is also very important in the regulation of calcium induced calcium release.
The simulations in previous research give contradictory
results~\cite{Cannell2013,Tanskanen2007}.  Hence, the explicit arrangement of
calcium channels in a cluster needs more investigations. 

The cluster size (the number of calcium channels in a cluster) is another important
index of the cluster, and was found to have a close relationship to the
diseases, such as Niemann-Pick type C1~\cite{Scott2021}. In Ref.~\cite{Qi2014},
the local calcium release from clusters with a few IP$_3$R channels was studied,
and nonlinearity was found for the interpuff interval and the first puff latency
against the inverse cluster size.  The effect of cluster sizes of both RyR's and
IP$_3$Rs was also studied in the
literature~\cite{Xie2019,Cao2013,Walker2014,Walker2015,Galice2018}. In the
recent years, more experiments were performed to study the frequency
distribution of cluster
sizes~\cite{Baddeley2009,Jayasinghe2018,Scott2021,Shen2019}.  Besides clusters
with large sizes, which have been observed in early experiments, many rogue RyR's,
as well as many small clusters, were also found in recent observations. It
has significance to study their roles in calcium signaling transduction. A recent
study suggests that  rogue RyR's greatly increase the initiation of Ca$^{2+}$
sparks, further contributing to the formation and propagation of Ca$^{2+}$
waves~\cite{Chena2018}.  More interestingly, recent super resolution imaging
shows a phenomenon that the frequency distribution  is exponential to the
cluster size, that is, it obeys the power
law~\cite{Baddeley2009,Jayasinghe2018}. It means that the clusters with one to
several RyR's constitute the majority of  all clusters.  

In the existing works, many studies focused on the dynamics in a cluster.
However, the effect of the spatial distribution of  calcium channels, such as the frequency distribution of  cluster sizes, are
beyond a cluster.  With the  development of super resolution imaging techniques,
more information about calcium channel distribution will be achieved in the future.
It is urgent to carry out systematic and appropriate simulation research beyond
the existing models to investigate irregular and random distributions of
channels, which is scarce in the literature.  If we go beyond the local cluster
and mean-field ansatz, a unified description of channel distribution at
different scales should be constructed.  In fact, the effect of channel
distribution on the calcium signaling transduction can be reflected by  Ca$^{2+}$
connections between calcium channels. If we know the effect of other channels on
the calcium concentration near the channel considered, the channel distribution
can be  described without loss of spatial information. The Ca$^{2+}$
connections between  calcium channels can be abstracted as the edges of a
network with the channels as nodes. The connection strength corresponds to a
weighted edge or the probability of connection by an unweighted edge.  

In the literature, there exist some attempts to study the calcium signal
transduction on the network of calcium channels. In
Refs.~\cite{Walker2014,Walker2015},  an adjacency matrix was introduced to
model calcium release in the heart. In fact, it equals  a network, but
only  the connections  between adjacent RyR's were considered. Such treatment can be
also found in Refs.~\cite{Iaparov2019,Iaparov2021} with more RyR's involved.
However, only simulation for a single cluster was performed to study the calcium
spark (blip). In Ref.~\cite{Hernandez-Hernandez2017}, with a simple two-state
model of the receptor, a bistable regime can emerge from the network dynamics
with several classical network architectures. By introducing the Keizer-Levine
model~\cite{Keizer1997} where a simplified mechanism mimicking adaptation has
been developed to reproduce experimental data from cardiac cells under a mean-field ansatz,  in our previous work the Ca$^{2+}$ induced Ca$^{2+}$ release was studied to explore the behavior of signaling networks with
multistates in which nodes are regulated by reaction rates nonlinearly~\cite{Jiang2021}. 

The studies suggest that different architectures of the network will affect the
bistable regime~\cite{Hernandez-Hernandez2017,Jiang2021}.  However, the calcium
oscillation was not reproduced in these models because only the state transition of calcium channel was considered. To reproduce the calcium oscillation,
more mechanisms, such as store leak and calcium pump, should be included to form
an open-cell model~\cite{Keizer1997}. Influx and efflux of Ca$^{2+}$ from an
external medium should be also considered to construct a more realistic model.
Under the mean-field ansatz, the Keizer-Levine model successfully generates the
calcium oscillation with these ingredients~\cite{Keizer1997}.  In this paper, we
will establish a theoretical frame to describe channel distribution by a
nonlinear network. The Keizer-Levine model will be extended to a network model to
include the effect of spatial distribution (Section~\ref{Mechanism}).  The
network with full connections, that is, complete graph, will be introduced to
check the model. 

With the established theoretical frame, three classical networks which reflect
different sides of spatial distribution of calcium channels will be introduced
to study their effect on calcium oscillation in Section~\ref{Network}. The
Erd\"os-R\'enyi network is introduced to check the equivalence of a
homogeneous network with the mean-field approximation. The heterogeneous
BaraB\'asi-Albert network is known for its scale-free characteristic~\cite{BA}.
It is introduced to reflect the power law of the frequency distribution of
cluster sizes if a cluster is taken as a node.  Another heterogeneous network,
a clique graph with nodes in a clique being fully connected, is also
introduced to simulate the compact clusters which have the same size, and
connect to other clusters weakly due to the large intercluster distance. In
Section~\ref{geometric network}, a more realistic model will be constructed
in a two-dimensional plane based on the experimental information about the internal
RyR arrangement of the cluster and the frequency distribution of cluster sizes. The
connections will be determined by calcium concentration gradient. With such
a geometric network, the role of clusters with small sizes will be investigated.
The discussion of the results and a summary  can be found in
Section~\ref{Summary}.

\section{Mechanism of Calcium Regulation}\label{Mechanism}

\subsection{General form} 

First, we present a general form  of a calcium regulation mechanism on a
network. The transition of a calcium channel, RyR or IP$_3$R, between its different
states is a kernel mechanism to reproduce the calcium oscillation. The calcium
channel has two basic states, open and closed. The Ca$^{2+}$ is released
from a channel in the open state. However, in a realistic model, there are often
more states required. Here, we assume that there exist $M$ states for a channel.
As in Ref.~\cite{Jiang2021}, we express the state as a vector $V=[0,\cdots,
1,\cdots,0]$ with one of its $M$ elements being 1 and others being 0.
$V_m=1$ means that the channel is in  state $m$.  For a system with $N$
channels, the total state of the system can be described by a matrix $R_{im}=V^i_m$,
with $i=1,2,\cdots N$ and $m=1,2,\cdots, M$.

The master equation for the state transition of calcium channel $i$ can be
generally written as, \begin{equation} \frac{d{\cal P}^i_{m}
(R_{im},t)}{dt}=\sum_{n}T^i_{mn}([{\rm Ca}^{2+}]_i)~ {\cal
P}^i_n(R_{in},t),\label{Eq: receptor} \end{equation} where ${\cal
P}^i_m(R_{im},t)$ is the probability of system in a state $R_{im}$ at a time
$t$. The transition rate for node $i$ between states $m$ and $n$ is described by
the element in $m$ row and $n$ column of a transition rate matrix $T^i([{\rm
Ca}^{2+}]_i)$, which is dependent on the calcium concentration  $[{\rm
Ca}^{2+}]_i$ at channel $i$. In different models, the transition rate matrix
$T^i([{\rm Ca}^{2+}]_i)$ can be different.

To determine the transition probability of a channel, the variation of calcium
concentration $[{\rm Ca}^{2+}]_i$ near this channel is required. The flux from
other open channels should be considered.  It will take some time before the
calcium ions released from an open channel diffuse to another channel.  In the
literature, the connection between an open channel and the channel considered
was often treated by the diffusion equation. However, the calcium signaling
propagates mainly with the calcium-induced-calcium-release mechanism, that is,
the Ca$^{2+}$ propagation between two neighbor channels is more important.
Besides, the diffusing Ca$^{2+}$ may attach into the buffers, such as
calmodulin, and no longer affect the calcium channels until deattaching. The
results in our previous work~\cite{Jiang2021b} and Ref.~\cite{Naraghi1997}
suggest that the Ca$^{2+}$ released from an open channel will disperse rapidly with
increasing distance.  Hence, the effect of calcium ions released from an open
channel is constrained in a small region.  As shown in
Refs.~\cite{Jayasinghe2018,Baddeley2009}, the cluster diameter and distance of
the nearest neighbor clusters have the largest distribution at values of 100~nm
and 200~nm, respectively, and the mean distance between RyR's in a cluster is
about 40~nm.  With the diffusion constant $D_{\rm Ca}=200\mu$m$^2$/s, we can
estimate the diffusion time as $t={r^2}/{2D_{\rm Ca}}$.  The above distances
correspond to times 0.025, 0.1, and 0.000004~ms, respectively. The diffusion
delay between the channels and clusters should be very small compared with the
mean RyR open time, about 2.2 ms~\cite{Za2003}. It was also discussed in
Refs.~\cite{Soeller1997,Valent2007,Iaparov2021} that the activation of RYR's
after an RYR opening proceeds much slower than the build-up of the calcium
gradient.  Hence, the diffusion delay between the RyR's around the open RyR can
be safely neglected, especially for the calcium oscillation considered here. In
the current paper, the  variation of calcium concentration $[{\rm Ca}^{2+}]_i$
near a channel is determined by the calcium gradient induced from open channels,
and the time delay is neglected.  

After neglecting the delay, the Ca$^{2+}$ connections between channels can be
further abstracted as edges of a complex network with calcium channels as nodes.
There exist two means to describe the  Ca$^{2+}$ connections, the weighted and
unweighted edges. The weight of a weighted edge can be determined by the calcium
concentration directly.  In the current paper, to introduce classical networks,
the unweighted networks will be adopted in calculation. In such networks, if a
channel pair has possibility $p$ to be connected by Ca$^{2+}$, the corresponding
element of adjacency matrix $A_{ij}$ has a possibility $p$ to be 1. For the
channel pair with no Ca$^{2+}$ connection, the element is set as 0. The treatment
will be explained more explicitly in Section~\ref{geometric network}. Obviously,
such a network is an undirected network.  Hence, for a channel $i$, the Ca$^{2+}$
flux received from other open channels can be written as \begin{equation}
J^R_i([{\rm Ca}^{2+}]_{j})=\sum_{jm}\gamma_j([{\rm Ca}^{2+}]_j)A_{ij}R_{jm}.
\label{Eq: fluxRyR} \end{equation} It is determined by release rate
$\gamma_j([{\rm Ca}^{2+}]_j)$ of all open channels $j$ which have connections to
the channel $i$, including $i$ itself if it is open. Such flux leads to the
increase of calcium concentration  near channel $i$.

In the realistic model,  more mechanisms, such as store leak, calcium pump, and
exchange with the external medium, should be included to form an open-cell model
where the oscillation can occur.  The total variation of the calcium
concentration near channel $i$  can be described as
\begin{equation} \frac{d[{\rm Ca}^{2+}]_i}{dt}=J^R_i([{\rm Ca}^{2+}]_{j})+J'_i([{\rm Ca}^{2+}]_i).\label{Eq: variation of calcium}
\end{equation}
The Ca$^{2+}$ flux from these mechanisms is denoted as $J'_i([{\rm Ca}^{2+}]_i)$. In a realistic model,  explicit positions about calcium pumps should be
introduced, for example, the distance between the calcium pump and calcium
channel. In the current paper, we focus on the mechanism of the state transition of calcium channels on network. The $J'_i([{\rm Ca}^{2+}]_i)$ is assumed
to be only dependent on the calcium concentration near the channel $[{\rm Ca}^{2+}]_i$ for
simplification.  

\subsection{Mean-field ansatz}

To understand the basic features of  network dynamics, we first provide the
approximate results under a simple spatially homogeneous mean-field ansatz. Under
this ansatz, all nodes, i.e. the calcium channels, are assumed to evolve with
time in the same manner.  If starting from a spatially homogeneous initial calcium
concentration, the equation for the variation of the calcium concentration in
Eq.~(\ref{Eq: variation of calcium}) should be the same for every node, that is,
independent on  $i$ and $j$. Therefore, the calcium concentration $[{\rm
Ca}^{2+}]_i$ should evolve homogeneously. The release rate $\gamma_j([{\rm
Ca}^{2+}]_j)$ is then deduced to $\gamma([{\rm Ca}^{2+}])$. The Ca$^{2+}$ flux
received from other open channels in Eq.~(\ref{Eq: fluxRyR}) can be rewritten as
 \begin{align}
  J^R_i=\gamma([{\rm Ca}^{2+}])\sum_{jm}A_{ij}R_{jm}.
\end{align} 
The state matrix $R_{jm}$ is deduced into
${R}_{1m}$ for all nodes, that is, all nodes have the same state $m$ at a time
point (here, we take state of node 1 as representative value). We reach
\begin{align}
\sum_{jm}A_{ij}R_{jm}=\sum_{j}A_{ij}\sum_m{R}_{1m}=\langle k\rangle \sum_{m=O}p_m.
\end{align}
Here we apply average degree $\langle
k\rangle=\frac{1}{N}\sum_{ij} A_{ij}=\sum_jA_{ij}$ and the faction of
channels in state $m$ is defined as $p_{m}=\frac{1}{N}\sum_i R_{im}={R}_{1m}$ due
to the spatial homogeneity under the mean-field ansatz. Since only the open channel can
release Ca$^{2+}$, only open states with $m=O$ are kept.   Equation
~(\ref{Eq: variation of calcium}) for the variation of calcium concentration
under the mean-field ansatz can be written as \begin{equation} \frac{d[{\rm
Ca}^{2+}]}{dt}=\gamma([{\rm Ca}^{2+}])\langle k\rangle \sum_{m=O}p_{m}+J'([{\rm
Ca}^{2+}]).\label{Eq: variation of calcium MF} \end{equation} The master
equation for the state transition of the calcium channel in Eq.~(\ref{Eq: receptor}) is
deduced to an equation of the fractions of channels in different states as,
\begin{equation} \frac{dp_{m}(t)}{dt}=\sum_{n}T_{mn}([{\rm
Ca}^{2+}])~p_{n}(t).\label{Eq: receptor MF} \end{equation}

To determine the explicit form of the flux $J^R([{\rm Ca}^{2+}])$, $J'([{\rm
Ca}^{2+}])$, and transition rate matrix $T([{\rm Ca}^{2+}])$, we compare
the general form of the mechanism under the mean-field ansatz in Eqs.~(\ref{Eq:
variation of calcium MF}) and (\ref{Eq: receptor MF}) with the Keizer-Levine model,
which is also under the mean-field ansatz~\cite{Keizer1997}. In the
Keizer-Levine model, four states of the RyR are introduced as $C_1$, $O_1$,
$C_2$, and $O_2$ to describe the experimental phenomena. The closed state $C_1$
is dominant at a low concentration $[{\rm Ca}^{2+}]$, for example, 0.1~$\mu$M.
If the $[{\rm Ca}^{2+}]$ increases, the RyR is activated from the closed state
$C_1$ to an open state $O_1$ at a rate of $k^+_a[{\rm Ca}^{2+}]^4$, and
deactivates back to $C_1$ state at a rate of $k^-_a$. As suggested in
Ref.~\cite{Keizer1997}, to keep the plateau, the open state $O_1$ also may be
activated to the second open state $O_2$  at a rate of $k^+_b[{\rm Ca}]^3$ and
back to the first open state $O_1$ at a rate of $k^-_b$. It is also important to
obtain the bistable regime~\cite{Jiang2021}.  To describe the adaption
phenomenon, the transition between the first open state $O_1$ and the second closed
state $C_2$ is also added.  The transition rates between these states are
described by the transition rate matrix $T([{\rm Ca}^{2+}])$ as listed in
Table~\ref{fmn}.  With the experimental data of Gy\"orke and
Fill~\cite{Gyorke1993}, the RyR kinetic constants $K^{\pm}_{(a,b,c)}$ have been
determined in Ref.~\cite{Keizer1997}, which are also given in  Table~\ref{fmn}.

\renewcommand\tabcolsep{0.27cm}
\renewcommand{\arraystretch}{2}
\begin{table}[h!]
  \caption{ Transition rate matrix $T([{\rm Ca}^{2+}])$ with parameters $k^+_a=1500~\mu$M$^{-4}$s$^{-1}$, $k^-_a=28.8$~s$^{-1}$, $k^+_b=1500~\mu$M$^{-3}$s$^{-1}$, $k^-_b=385.9$~s$^{-1}$, $k^+_c=1.75$~s$^{-1}$, and $k^-_c=0.1$~s$^{-1}$~\cite{Keizer1997}.\label{fmn}}
  \centering
  \begin{tabular}{c | cccccc  }\bottomrule[1pt]
    State & $C_1$                        & $O_1$                                   & $C_2$    & $O_2$    \\\hline
    $C_1$ & $-k_a^+ [{\rm Ca}^{2+}]^4$ & $k^-_a$                                 & $--$     & $--$     \\
    $O_1$ & $k_a^+ [{\rm Ca}^{2+}]^4$  & $-k^-_a-k_b^+[{\rm Ca}^{2+}]^3-k^+_c$ & $k^-_c$  & $k^-_b$  \\
    $C_2$ & $--$                         & $k^+_c$                                 & $-k^-_c$ & $--$     \\
    $O_2$ & $--$                         & $k_b^+[{\rm Ca}^{2+}]^3$              & $--$     & $-k^-_b$ \\
    \hline
    \bottomrule[1pt]
  \end{tabular}
\end{table}

By comparing the Ca$^{2+}$ flux from the RyR's in the Keizer-Levine
model~\cite{Keizer1997} and that under the mean-field ansatz in Eq.~(\ref{Eq:
variation of calcium MF}), the transition rate of channels can be determined as,
\begin{align} \gamma([{\rm Ca}^{2+}]) & =f\nu'_1 ([{\rm Ca}^{2+}_s]-[{\rm
Ca}^{2+}]),\label{Eq: rate} \end{align} where a factor $f=0.01$ is introduced as
the fraction of Ca$^{2+}$ that is free in the cytoplasm~\cite{Tse1994}. The
$\nu'_1$ is the release rate constant for the RyRs. We would like to emphasize
that the definitions of release rate under the mean-field ansatz and for a
single channel  should be different.  The $\nu'$ is realistic release rate
constant which reflects the ability of a channel receptor. The
$\nu_1$ adopted under the mean-field ansatz is, in fact, an effective quantity, and
dependent on the average degree $\langle k\rangle$.  In the current work, we
relate the release rate for a channel to that under the mean-field ansatz as
$\nu'_1=\nu_1/\langle k\rangle$.  The $\nu_1$ was determined as 40 s$^{-1}$
based on Friel's analysis of the bullfrog sympathetic neuron~\cite{Friel1995}.
With such relation, the mean-field approximation of a network will be deduced to
the Keizer-Levine model. 

As in Eq.~(\ref{Eq: rate}), the release of calcium from the RyR's is assumed to be proportional to the
concentration gap between cytoplasm and store, i.~e., $[{\rm Ca}^{2+}]$ and
$[{\rm Ca}^{2+}_s]$.  The calcium concentration of store is assumed as
$[{\rm Ca}^{2+}_s]=(C_0-[{\rm Ca}^{2+}])/c_1$. The factor $c_1$ is sometimes
referred to as the ratio of effective volume of the store to the cytoplasm, and is
determined as 0.15 based on the experimental data in Ref.~\cite{Alberts1989}.
The $C_0$ is the total free-Ca$^{2+}$ concentration in the cell. If the cell is
open to external medium, it should not be fixed, but  vary with influx
and efflux from the Plasma membrane calcium (PMCA)  pump according to the equation~\cite{Keizer1997}, 
\begin{align} 
  \frac{dC_0}{dt} & =f\left(j_{\rm in}-\nu_{\rm out}\frac{[{\rm
Ca}^{2+}]^2}{[{\rm Ca}^{2+}]^2+K^2_{\rm out}}\right),\label{Eq: C0} 
\end{align} 
with $j_{\rm in}$ being the influx rate, which has a value $j_{\rm
in}=1~\mu$Ms$^{-1}$ obtained from a current of 0.1~pA in a cell with a volume of
1000~$\mu$m$^{3}$.  $\nu_{\rm out}$ and $K_{\rm out}$ are the maximal rate and
dissociation constant of the PMCA pump, and  chosen as 9.0~$\mu$Ms$^{-1}$ and
0.6~$\mu$Ms, respectively~\cite{Carafoli1994,Tepikin1992,Keizer1997}.

Besides the exchange of Ca$^{2+}$ with external medium, the Ca$^{2+}$ exchange
with the calcium store should be included. Hence, the flux $J' [{\rm Ca}^{2+}]$ can be
described as~\cite{Keizer1997},
\begin{align} J' [{\rm Ca}^{2+}] & =f\left(\nu_2
([{\rm Ca}^{2+}_s]-[{\rm Ca}^{2+}])-\nu_3\frac{[{\rm Ca}^{2+}]^2}{[{\rm
Ca}^{2+}]^2+K^2_3}\right.\nonumber \\ & \left. +j_{\rm in} -\nu_{\rm
out}\frac{[{\rm Ca}^{2+}]^2}{[{\rm Ca}^{2+}]^2+K^2_{\rm out}}\right).
\end{align} 
The first and second terms are for the calcium leak  from the calcium store, and
returning of ${\rm Ca}^{2+}$ to calcium store through the SERCA bump,
respectively. The third and fourth terms are the influx and efflux of ${\rm
Ca}^{2+}$ exchange with the external medium, respectively, as in Eq.~(\ref{Eq: C0}).
Here the dissociation constant of SERCA pump $K_3=0.3~\mu$M~\cite{Lytton1992}.
The $\nu_2$  and $\nu_3$ are the rate constants for store leak and the SERCA
pump, and chosen as 0.1~$\mu$Ms$^{-1}$ and 120~$\mu$Ms$^{-1}$,
respectively~\cite{Keizer1997}.

\begin{figure}[h!]
  \includegraphics[bb=0 0 600 820,scale=0.41,clip]{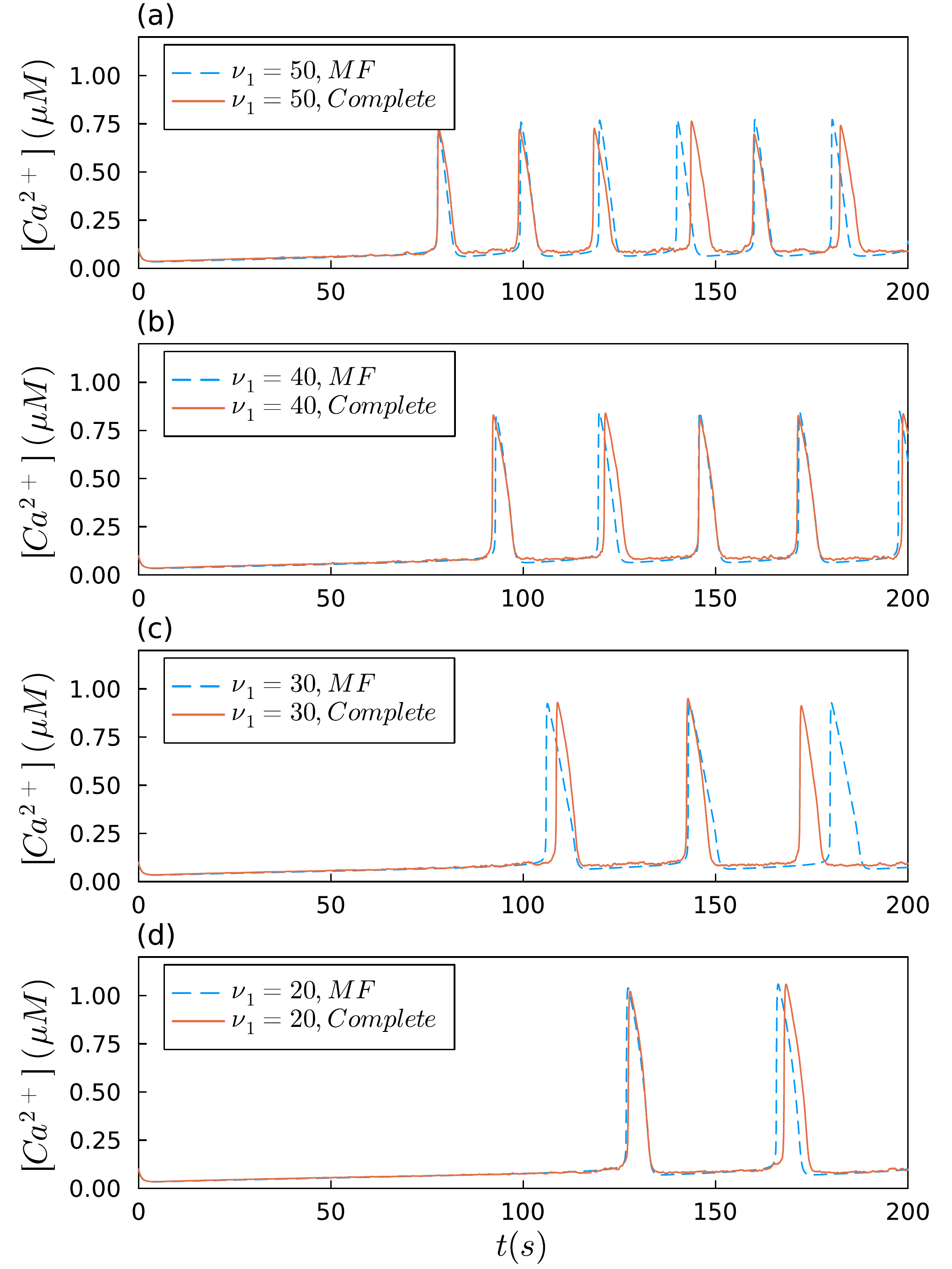}
\caption{Calcium oscillation in mean-field ansatz (MF) and on the complete graph network with different release rate constant of RyR's  under mean field ansatz as $\nu_1=50$, 40, 30, and 20~s$^{-1}$ in  (a)-(d), respectively.}
\label{Fig: MF}
\end{figure}

The calcium oscillation produced with a set of the parameters under the
mean-field ansatz is presented in Fig.~\ref{Fig: MF}. 
Different from the
original Keizer-Levine model, in  our calculation, Eqs.~(\ref{Eq: variation of
calcium MF}) and (\ref{Eq: receptor MF}) are adopted directly without separation
between fast and slow time scales.  As discussed above, in Eq.~(\ref{Eq: rate}),
the release rate of RyR's under the mean-field ansatz can be related to the
average degree of the network.  Here, we present the results with different
release rates under the mean-field ansatz. As shown in Fig.~\ref{Fig: MF}, the calcium
oscillation can be reproduced  with the $v_1$ considered, and very close to the
results with separation between fast and slow time scales~\cite{Keizer1997}. One
can find that larger release rates result in the calcium oscillation with a smaller
interspike interval and smaller amplitude.  With the increase of release
rate $\nu_1$ from 20 to 50~s$^{-1}$, relaxation time of the oscillation becomes small.

\subsection{Calcium oscillation on complete graph network}

The complete graph with full connections between nodes is introduced here
to check the network model.  Obviously, such a network is homogeneous and only
dependent on the number of nodes.  Since the calcium concentration near a
channel will increase with the total number of channels,  the complete graph is only
appropriate to describe a small  system, such as a compact cluster, but can not be
applied to the channel distribution at a large scale.   However, it can be well
described by the mean-field approximation, and used to check the network
formalism adopted in the current paper. The results on this network should well
fit well with the mean-field approximation with $\nu'_1=\nu_1/(N-1)$ for large channel number
$N$, which is chosen as 1000 in the current paper. Under the mean-field ansatz,
the $[{\rm Ca}^{2+}]_i$, $\gamma_j([{\rm Ca}^{2+}]_j)$, and $J'_i([{\rm
Ca}^{2+}]_i)$ are the same for all channels. After the network is introduced,
these quantities should be set differently for every channel. After inserting them
into Eqs.~(\ref{Eq: receptor})-(\ref{Eq: variation of calcium}), one can reach a
formalism for the calcium signaling on a network.

In the current paper, we adopt the \texttt{Julia} language to perform simulations.
The Gillespie algorithm is employed to  deal with the transition of RyR states.
There exist $n_r=6$ types of transitions between different states of every
channel as shown in Table~\ref{fmn}.  Hence, at a certain time point, total
$n=N n_r=6000$ events may occur; the possibility for each event can be
obtained with the transition rate and propensity as $\alpha_i$ with $i$ from 1
to $n$. As usual with the Gillespie algorithm~\cite{Gillespie1977}, two random numbers
$r_1$ and $r_2$ are generated. One of them, $r_2$ is used to judge which event
will occur. If $\sum_{i=1}^{j-1}\le r_2\alpha\le\sum_{i=1}^j$ with total
possibility $\alpha=\sum_{i=1}^{n}\alpha_i$, the $i{\rm th}$ event occurs.
Another $r_1$ is used to generate the time interval until the next event as $\Delta
t=-\log r_1/\alpha$. The calcium concentration $[{\rm Ca}^{2+}]_i$ and $C_{0i}$
also have variations $\frac{d[{\rm Ca}^{2+}]_i}{dt}|_{t}\Delta t$ and
$\frac{dC_{0i}}{dt}|_{t}\Delta t$.  With such an algorithm, the calcium
concentration and the state of each RyR evolve with the time $t$. 

The results with the complete graph are also presented in  Fig.~\ref{Fig: MF}.
As shown in Eq.~(\ref{Eq: rate}), to compare the results with these in the
mean-field approximation, the release rate $\nu_1$ is divided by the average
degree $\langle k\rangle$. Under such definition, the model with complete graphs
corresponds to the mean-field approximation with the same value of $\nu_1$. The
global calcium concentration is defined as the average of all RyRs, $[{\rm
Ca}^{2+}]=\sum_{i=1}^N[{\rm Ca}^{2+}]_i/N$.  As expected, the result shows that
simulation of global calcium concentration with complete graphs produces almost
the same interspike time and amplitude as the mean-field approximation.  The
interspike time and amplitude with the mean-field approximation are identical
for each period while little deviation can be found for the simulation with the
network due to the uncertainties of the random numbers introduced in the
Gillespie algorithm.  Decrease of the interspike interval and amplitude can be
found with the increase of the release rate $\nu_1$.

\section{Calcium oscillation on classical networks}\label{Network}

The complete graph can be related to a compact cluster, in which all channels
are connected to each other. The above calculation suggests that such a structure
can be well described by the mean-field approximation as in
Ref.~\cite{Iaparov2019}.  In the current paper, the spatial distribution of the
calcium channels beyond the cluster is considered, which is very complex. The
intracellular space is crowded with various obstructions and
organelles~\cite{Rothman1994}.  The diffusing Ca$^{2+}$ has to navigate between
these complex intracellular structures. Most of the calcium channels distribute
on the smooth endoplasmic reticulum with a complex three-dimensional network
structure~\cite{Shen2019}.  Even for the rough endoplasmic reticulum with a flat
layer shape, its folding structure in the cytoplasm makes some parts of
different sheets very close to each other~\cite{Shibata2006}.  Ca$^{2+}$ can be communicated between distant regions of endoplasmic reticulum.
Combined with the  cluster structure, the connections between calcium
channels should be very complex, and suitable to be described as a complex
network. 

With the theoretical frame constructed in the above section, in this section,
three classic networks will be introduced to study the effect of different
network architectures  on the calcium oscillation. In this section, the
homogeneous Erd\"os-R\'enyi network,  heterogeneous BaraB\'asi-Albert, and
clique graph networks~\cite{Newman} will be considered to reflect the
characteristics of spatial distribution of calcium channels. The procedure to
study these networks is almost the same as the complete graph after the
adjacency matrix $A$ is replaced.  In the following simulations, the networks
and its adjacency matrix will be  generated with the \texttt{Graphs} package of
the \texttt{Julia} Lagrangian.  And the diagonal elements will be set 1 to include the effect the channel itself.

\subsection{Erd\"os-R\'enyi network}

\begin{figure}[h!]
  \includegraphics[bb=0 0 820 310,scale=0.41,clip]{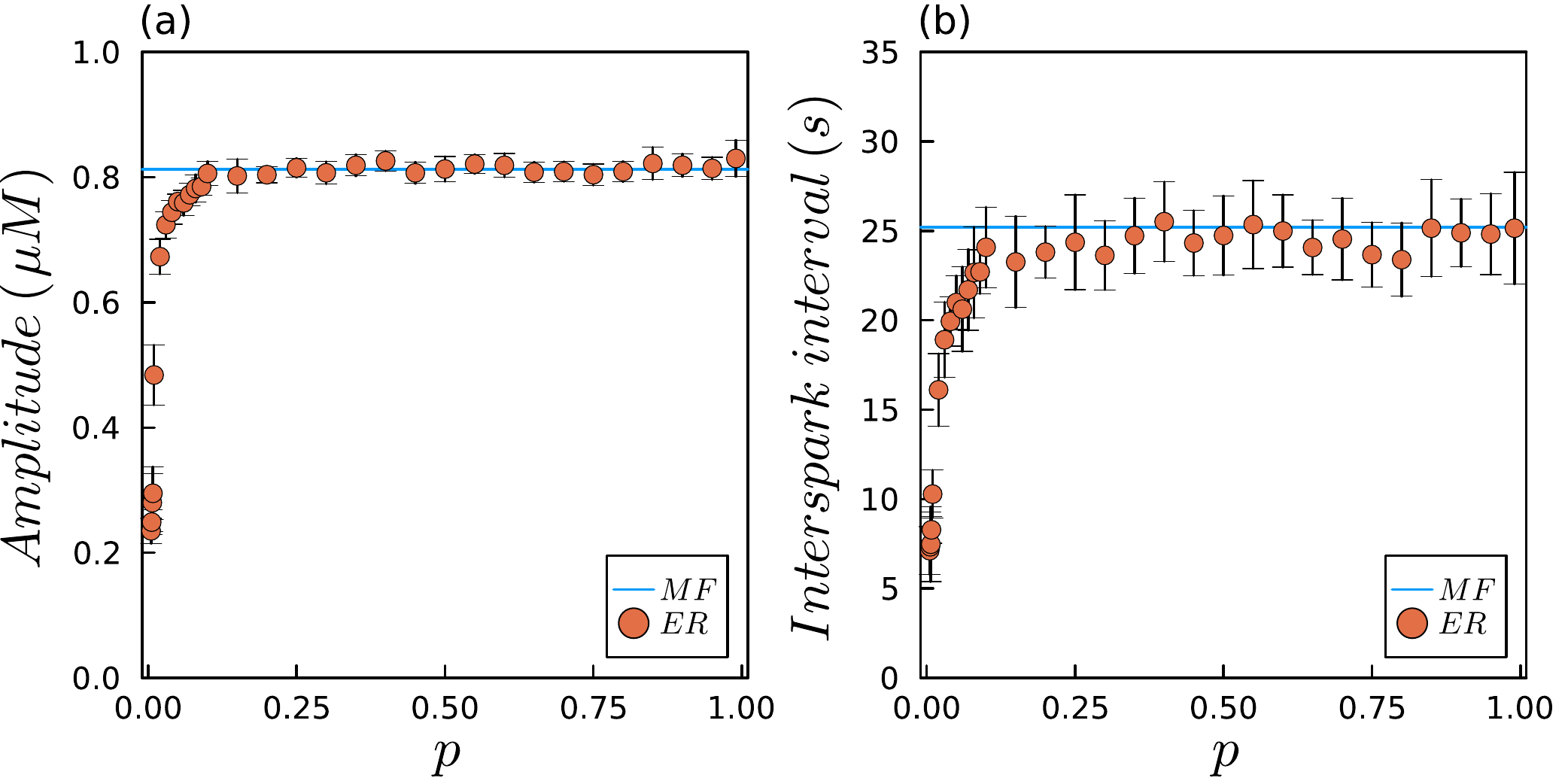}
\caption{Average amplitude and interspike interval of calcium oscillation on the Erd\"os-R\'enyi (ER) network with the variation of possibility $p$. The values in the  mean-field approximation are given as horizontal lines. Other parameters are chosen as those in Fig.~\ref{Fig: MF} with $\nu_1=40$~s$^{-1}$, and $N=1000$.}
\label{Fig: ISI_ER}
\end{figure}

The Erd\"os-R\'enyi network with $N$ nodes, denoted as $G_{ER}(N,p)$, is
generated by adding edges into the node pairs with a possibility $p$. It is
homogeneous because every node in the network has equal status. The
Erd\"os-R\'enyi network can describe the characteristic of channel distribution
at large scale where the mechanism of calcium regulation can be successfully
described  in the mean-field approximation.  Different from the complete
graph, the  Erd\"os-R\'enyi network is not completely connected and randomness
is introduced. As a homogeneous network such as the complete graph, the behavior of
evolution of calcium concentration should also be analogous to those in the mean-field approximation. In Fig.~\ref{Fig: ISI_ER}, an average amplitude of ten spikes
and average interspike interval of ten periods with the variation of possibility
$p$ are presented with the standard deviations.

In Fig.~\ref{Fig: ISI_ER}, one can find that in a large range of possibility $p$
larger than 0.1, the amplitude is about 0.8~$\mu$M and the interspike interval
is about 25~s, which fit the  mean-field approximation as expected.  However,
if the possibility $p$ becomes very small, both amplitude and interspike
interval decrease rapidly and the oscillation disappears rapidly. In the current
paper, we fix the release rate constant $\nu_1$ in the mean-field approximation
at 40~s$^{-1}$, and the rate for the network can be obtained as
$\nu'_1=\nu_1/\langle k\rangle$.   For the Erd\"os-R\'enyi network, the average
degree $\langle k\rangle=p(N-1)$, which means that the large possibility $p$
leads to small release rate constant $\nu'_1$ for calcium channels, which are
nodes of the network.  Hence,  on a denser network, smaller ability of transporting
calcium ions is required.

\begin{figure}[h!]
  \begin{center}
    \begin{overpic}[bb=0 0 3000 930,scale=0.365,clip]{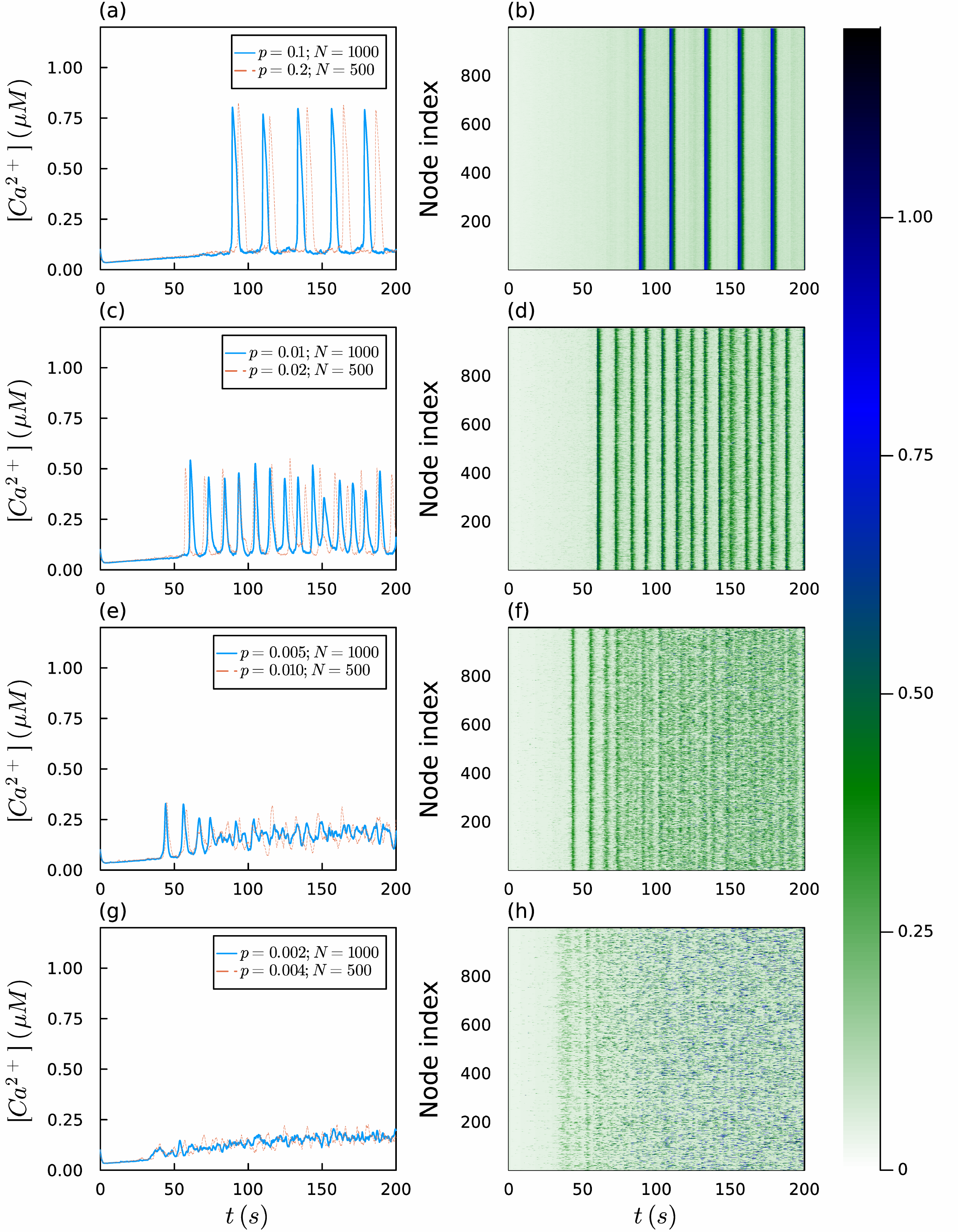}
      \put(2.4,25.6){\includegraphics[scale=.007]{ER_1000_01_0.pdf}}
      \put(2.4,18.2){\includegraphics[scale=.007]{ER_1000_001_0.pdf}}
      \put(2.4,10.8){\includegraphics[scale=.007]{ER_1000_0005_0.pdf}}
      \put(2.4,3.40){\includegraphics[scale=.007]{ER_1000_0002_0.pdf}}
    \end{overpic}
  \end{center}
\caption{Evolution of calcium concentration $[{\rm Ca}^{2+}]$ with the variation of time $t$ on the Erd\"os-R\'enyi network. Global calcium concentrations with $p$=0.1, 0.01, 0.005, 0.002 and $N$=1000 (full blue line), with $p$=0.2, 0.02, 0.01, 0.004 and $N$=500  (dashed red line)  in  (a), (c), (e), (g), respectively.  Calcium concentrations  for each node with $p$=0.1, 0.01, 0.005, 0.002 and $N=1000$   in (b), (d), (f), (h), respectively.  The networks with different $p$ and $N=1000$ are also visualized, and presented in corresponding (a), (c), (e), (g). Other parameters are chosen as those in Fig.~\ref{Fig: ISI_ER}.} \label{Fig: ER}
\end{figure}

To provide a more explicit picture about the oscillation on the network, in  Fig.~\ref{Fig: ER}, the evolution of calcium concentration $[{\rm
Ca}^{2+}]$ on the Erd\"os-R\'enyi network is presented. The calcium
concentration $[{\rm Ca}^{2+}]_i$ for every node, i.~e.  near every channel, is
also illustrated. 

Since the dense networks with large $p$ values show similar behaviors as
suggested in Fig.~\ref{Fig: ISI_ER}, only some selected small $p$ values are
considered.  As expected, on the network $G_{\rm ER}(1000, 0.1)$,  the evolution of
global calcium concentration with time $t$ is similar to these in the mean-field approximation in Fig.~\ref{Fig: MF}(b), which can be digitalized as amplitude
and interspike interval close to the values in mean-field approximation in
Fig.~\ref{Fig: ISI_ER}.  The evolution of calcium concentration with time for
each channel exhibits almost the same behavior as shown in Fig.~\ref{Fig:
ER}(b), which indicates the synchronization of the system, and results in a
wonderful global oscillation.

Increase of sparsity will reduce the homogeneity of network.  On the network
$G_{\rm ER}(1000, 0.01)$, the oscillation is still reproduced well  with a smaller
interspike interval and amplitude, which deviate from the mean-field approximation.  Moreover, a few nodes do not behave within the same manner as
the majority, which makes the fringes fluffy as shown in Fig.~\ref{Fig: ER}(d). It
reflects the appearance of desynchronization. With the decrease of the $p$
value, the network becomes sparser, and the synchronization of nodes is broken
down gradually.  On the network $G_{\rm ER}(1000, 0.005)$, more nodes are out of
synchronization, which makes the oscillation of the global calcium concentration
unobvious.  On the network $G_{\rm ER}(1000, 0.002)$ where some nodes are even not
connected, the synchronization is almost broken down totally, and the global
oscillation of calcium concentration disappears also as shown in Fig.~\ref{Fig:
ER}(h). 

We  make a simple check about the effect of node number of the system.
The results are presented in Figs.~\ref{Fig: ISI_ER}(a),~\ref{Fig: ISI_ER}(c),~\ref{Fig: ISI_ER}(e), and~\ref{Fig: ISI_ER}(f) as dashed red
lines. In these simulations, only 500 nodes are considered, which is a half
of the node number in the above simulation. At the same time the connection
possibility $p$ is doubled to keep the average degree unchanged. Under such
treatment, the size of network is reduced, but the structure is not changed.
The results suggest that the calcium oscillations exhibit almost the same
behavior in two cases.

\subsection{BaraB\'asi-Albert network}

\begin{figure}[h!]
  \includegraphics[bb=0 0 820 310,scale=0.41,clip]{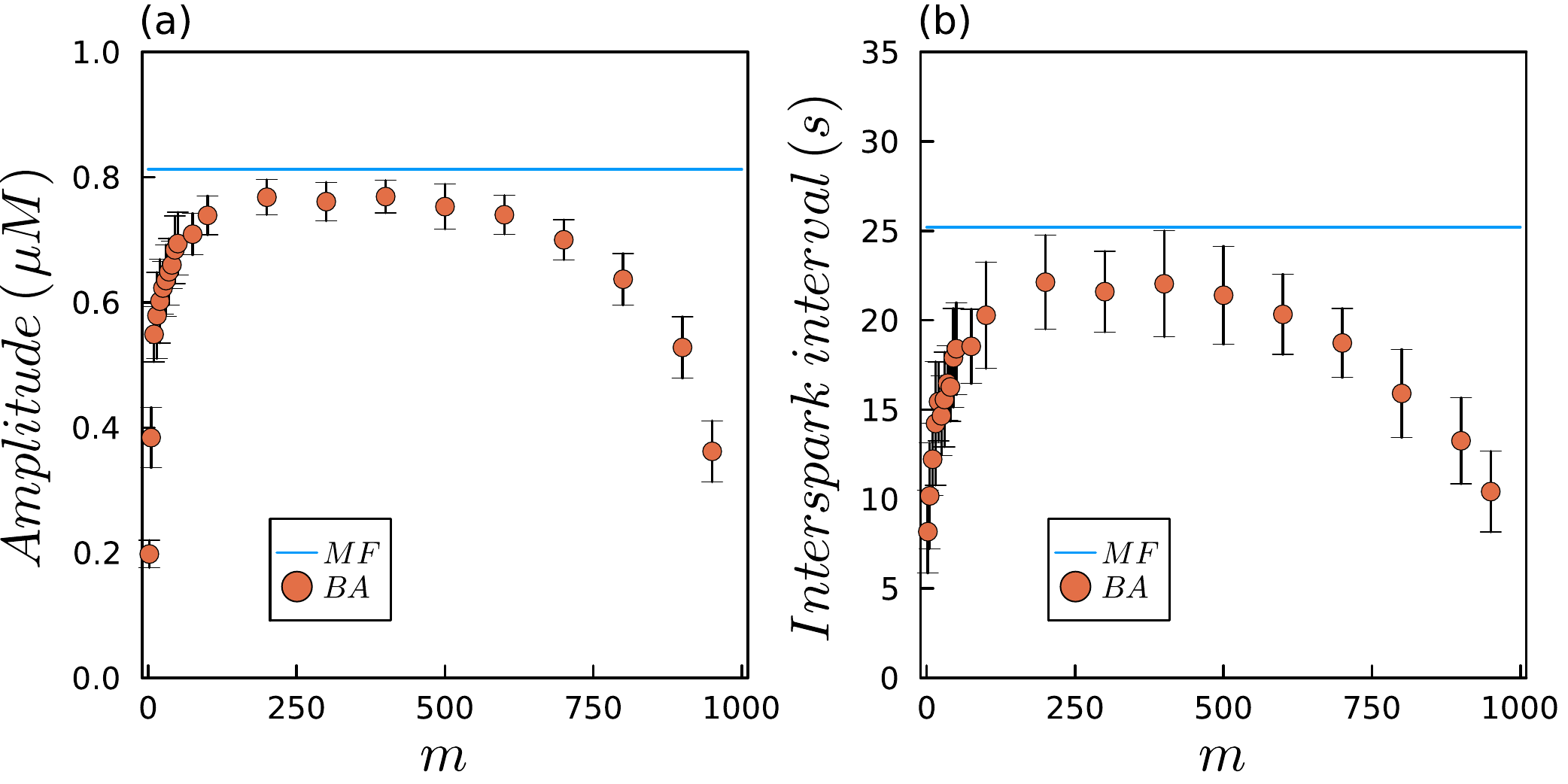}
  \caption{Average amplitude and interspike interval  of calcium oscillations on the BaraB\'asi-Albert (BA) network with the variation of parameters $m$. The values in mean-field approximation are given as horizontal lines. Other parameters are chosen as those in Fig.~\ref{Fig: ISI_ER}.}
  \label{Fig: ISI_BA}
\end{figure}

Different from the Erd\"os-R\'enyi network, the BaraB\'asi-Albert network is
heterogeneous. The BaraB\'asi-Albert network is generated by adding a sequence
of $m$ new nodes to an existing network, so can be denoted as
$G_{\rm BA}(N,m)$~\cite{BA}.  As a scale-free network, hub nodes will emerge and
play a more important role in the network dynamics. Recent super resolution
nanoscale imaging shows that the cluster sizes satisfy the power
law~\cite{Baddeley2009,Jayasinghe2018}. If we take a cluster as a node and
assume that the connection ability of a cluster to other clusters is
proportional to its size, the network should be a scale-free network.  Here we
adopt the BaraB\'asi-Albert  network to reflect sucha  characteristic of channel 
distribution.  In Fig.~\ref{Fig: ISI_BA}, the average amplitude and average
interspike interval of calcium oscillations  are presented. 

In the case of Erd\"os-R\'enyi network, the simulation fits the mean-field
results very well with large  $p$ values.  The BaraB\'asi-Albert network
exhibits a quite different behavior with large $m$ values where the network is
also dense. The maximum can be found at $m$ values of several hundreds, where
the large average degrees of the network can  also be found. Both amplitude and
interspike intervals are close but still below the mean-field results. No stable
range such as the Erd\"os-R\'enyi network can be found in the case of the
BaraB\'asi-Albert network. The interspike interval and amplitude vary
with the variation of the $m$ value. If the $m$ value increases further, both
amplitude and interspike interval decrease with the decrease of network density,
and deviate further from the mean-field results.   

\begin{figure}[h!]
  \begin{center}
    \begin{overpic}[bb=0 5 3000 930,scale=0.365,clip]{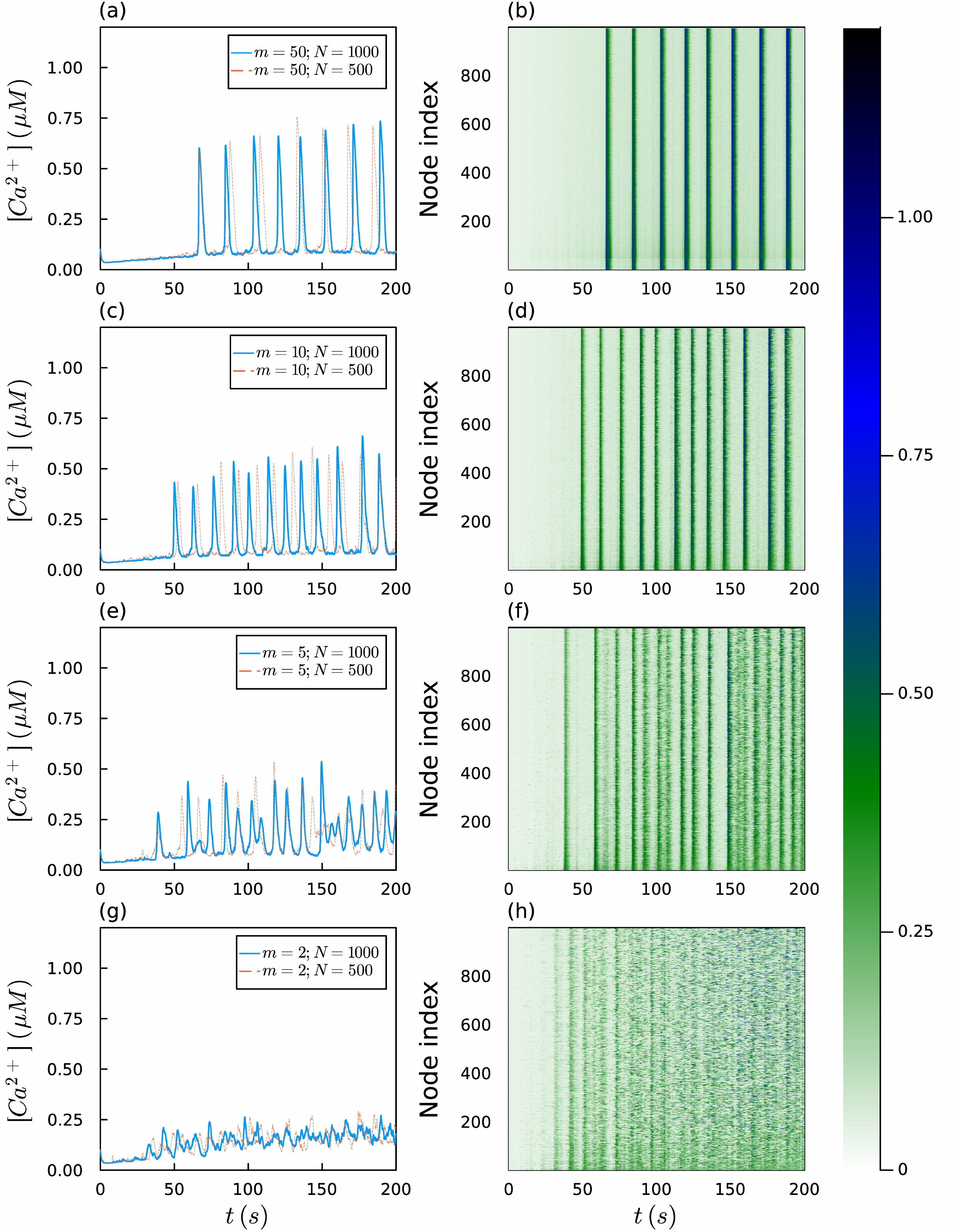}
      \put(2.4,25.6){\includegraphics[scale=.007]{BA_1000_50_0.pdf}}
      \put(2.4,18.2){\includegraphics[scale=.007]{BA_1000_10_0.pdf}}
      \put(2.4,10.8){\includegraphics[scale=.007]{BA_1000_5_0.pdf}}
      \put(2.4,3.40){\includegraphics[scale=.007]{BA_1000_2_0.pdf}}
    \end{overpic}
  \end{center}
  \caption{Evolution of calcium concentration $[{\rm Ca}^{2+}]$ with the variation of time $t$ on the BaraB\'asi-Albert network. Global calcium concentrations with $m$=50, 10, 5, 2 and $N=1000$ (full blue line) or $N=500$  (dashed red line)  in (a), (c), (e), (g), respectively.  Calcium concentrations  for each node with $m$=50, 10, 5, 2 and $N=1000$  in  (b), (d), (f), (h), respectively.  The networks with different $m$ and $N=1000$ are also visualized, and presented in corresponding  (a), (c), (e), (g). Other parameters are chosen as those in Fig.~\ref{Fig: ISI_ER}.}
  \label{Fig: BA}
\end{figure}

For smaller $m$, a rapid decrease of the amplitude and  interspike interval are
found, as in the Erd\"os-R\'enyi network. To provide more explicit details, the
evolution of the calcium concentration $[{\rm Ca}^{2+}]$ with the variation of
time $t$ on the BaraB\'asi-Albert network is presented in Fig.~\ref{Fig: BA}.
The calcium concentration $[{\rm Ca}^{2+}]_i$ for every node is also illustrated.  

Since the BaraB\'asi-Albert network with small $m$ value is more important, here, we
perform the simulations with $m$=50, 10, 5, 2 and $N$=1000. For the network
$G_{\rm BA}(1000, 50)$, the average degree is 95, which is almost the same as the
network $G_{\rm ER}(1000,0.1)$. The simulation on the network $G_{\rm ER}(1000,0.1)$ is
almost the same as the mean-field result, while a large deviation appears in the
simulation on network $G_{\rm BA}(1000, 50)$. Obviously, a small amplitude and high
frequency can be found, which suggests that the heterogeneous structure seriously affects
the calcium oscillation. However, good  synchronization is still
observed as in the network $G_{\rm ER}(1000,0.1)$ with clear fringes as shown in
Fig.~\ref{Fig: BA}(b). Most nodes behave in the same manner with the variation
of time, which results in a wonderful global oscillation, as shown in
Fig.~\ref{Fig: ER}(a). 

Decreasing the $m$ value will reduce the average degree  of the network, which means
an increase of global sparsity. With $m$=10 and 5, the oscillation is still
reproduced with a smaller interspike interval and smaller amplitude, which deviate
further from the results in the  mean-field approximation. The desynchronization
also appears, which leads to stochastic disturbance on the global oscillation
and fluffy on  the fringes, as shown in Fig.~\ref{Fig: ER}(d) and ~\ref{Fig: ER}(f).  With $m$=2,
some nodes are not even  connected, the synchronization is almost broken down
totally, and the oscillation of global calcium concentration also disappears, as
shown in Fig.~\ref{Fig: BA}(h). 

The network with small size is also introduced to test the effect of node number. The results are presented in Fig.~\ref{Fig: ISI_BA}(a), ~\ref{Fig: ISI_BA}(c), ~\ref{Fig: ISI_BA}(e), and~\ref{Fig: ISI_BA}(f) as
dashed red lines. Different from the Erd\"os-R\'enyi network, the
BaraB\'asi-Albert network with the same $m$ value has the same structure except the
scale  of the network. Here,  500 nodes are considered for checking. The
results suggest that the calcium oscillations exhibit almost the same behaviors
in two cases.

\subsection{Clique graph}

\begin{figure}[h!]
  \begin{center}
    \begin{overpic}[bb=0 5 3000 930,scale=0.365,clip]{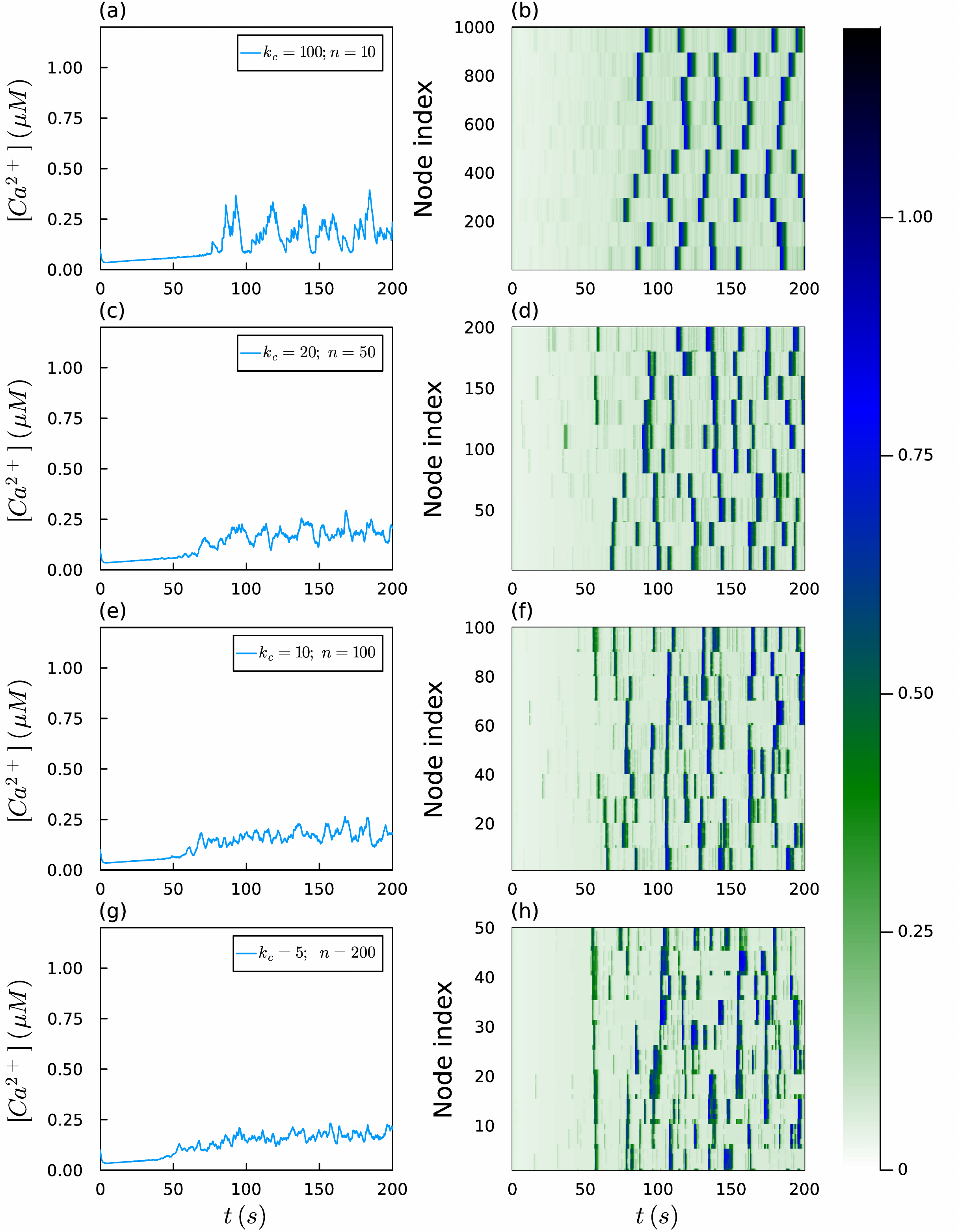}
      \put(2.4,25.2){\includegraphics[scale=.007]{Clique_1000_100_10.pdf}}
      \put(2.4,17.8){\includegraphics[scale=.007]{Clique_1000_20_50.pdf}}
      \put(2.4,10.4){\includegraphics[scale=.007]{Clique_1000_10_100.pdf}}
      \put(2.4,3.00){\includegraphics[scale=.007]{Clique_1000_5_200.pdf}}
    \end{overpic}
  \end{center}
  \caption{Evolution of calcium concentration $[{\rm Ca}^{2+}]$ with the variation of time $t$ on the clique graph network. Global calcium concentrations with $k_c$=100, 20, 10, 5 are in  (a), (c), (e), (g), respectively.  Calcium concentrations  for each node with $k_c$=50, 10, 5, 2 and $N$=1000  are in (b), (d), (f), (h), respectively.  The networks with different $k_c$ values  are also visualized, and presented in corresponding  (a), (c), (e), (g). Other parameters are chosen as those in Fig.~\ref{Fig: ISI_ER}.}
  \label{Fig: clique}
\end{figure}

In the literature, many studies were performed locally under the assumption that the
cluster is compact and has a regular shape~\cite{Rudiger2010,Qi2014}. In such
a picture, in a cluster, an open channel affects all other channels through calcium
releasing. However, the calcium connections between clusters should be weak
because the distances between the clusters are assumed to be large. It
corresponds to a clique graph,  in which several nodes are completely connected,
while there is at most one connection between two clusters. The clique graph
$G_{\rm cl}(k_c,n)$ consists of $n$ completely connected cliques with $k_c$ nodes.
In Fig.~\ref{Fig: clique}, we choose four architectures with $N$=1000 nodes and
$k_c$= 100, 20, 10, 5.

On network $G_{\rm cl}(10,100)$, 1000 channels are divided into ten clusters with 100
channels. The results suggest that in clusters  local oscillation is
produced due to  local synchronization as shown in Fig.~\ref{Fig: clique}(b).
However, weak connections between the clusters make the synchronization between
clusters impossible, as shown in Fig.~\ref{Fig: clique}(a). Without global
synchronization, the local oscillations in clusters can not merge into a global
oscillation. Such results suggest that, if the edges of cluster are distinct and
the distances between neighbor clusters are too large, the lack of global
synchronization will prevent global oscillation.  

The current results also contain the information of  local dynamics of 
clusters with different sizes. A large cluster  can produce local oscillation
with local synchronization. If the size of a cluster is reduced to ten channels,
local synchronization is still kept in a cluster. However, the oscillation
obviously becomes  aperiodic, and the interspike interval obviously varies  with
the time evolution. If cluster size  becomes very small, local synchronization
is even broken down in a cluster due to the connections with other clusters
as shown in Fig.~\ref{Fig: clique}(h).

\section{Calcium oscillation on geometric network with cluster structure}
\label{geometric network}

In the previous section, the calcium oscillations on three classic networks are
studied.  The oscillation can be produced on the homogeneous Erd\"os-R\'enyi
network with a large enough degree and coincides with the mean-field
approximation.  On the heterogeneous BaraB\'asi-Albert network, the calcium
oscillation can happen, however, it deviates from the mean-field approximation with
most of the parameter values. The synchronization is found to be important in these
two networks. For a sparse network, the oscillation disappears mainly due to the
desynchronization.  The results on the clique graph network show the importance
of the synchronization more clearly. In such a network, the oscillation can be
found in every cluster if its size is large enough. However, the connections
between the clusters are very weak, which makes the  synchronization between the
clusters hard to maintain. 

The cluster structure has been confirmed by a large amount of experiments.  To
recover the synchronization, the connections between the clusters should be
enhanced, which means that the distances between neighbor clusters should be
shortened, or more clusters should be added between the existing clusters.
However, in the above simulations, the classical networks are introduced
directly. The geometric characteristics of the cluster are not included. It
makes the understandings of the effect of clusters difficult. In this section,
we will introduce a more realistic model to study the calcium oscillation. As
said above, the real spatial distribution of calcium channels is very complex.
Here, we consider a simple but very important picture, in which all calcium
channels distribute on a two-dimensional plane with a cluster structure. In a cell,
such spatial distribution can be found in the rough endoplasmic reticulum with
a flat-layer shape~\cite{Shibata2006}. In the literature, there exists much
information about the cluster structure, which will be adopted to constrain the
model to construct a more realistic network for the spatial distribution of
calcium channels.

\subsection{Construction of geometric network}

The experimental observations suggest that the clusters with random non-compact
irregular shapes scatter randomly on the endoplasmic
reticulum~\cite{Baddeley2009,Jayasinghe2018,Shen2019}. Hence, the spatial
distribution of calcium channels should be generated randomly and converted to a
network based on experimental information. The geometric network will be
constructed in the following steps: (1) randomly generate a series of cluster
sizes, (2) randomly generate the clusters with
different sizes, (3) randomly distribute the clusters in a two-dimensional plane,
and (4) convert the spatial distribution of the RyR's into a geometric network.  In the
following, the explicit method will be presented.

\begin{figure}[h!]
  \includegraphics[bb=0 0 1200 320,scale=0.41 ,clip]{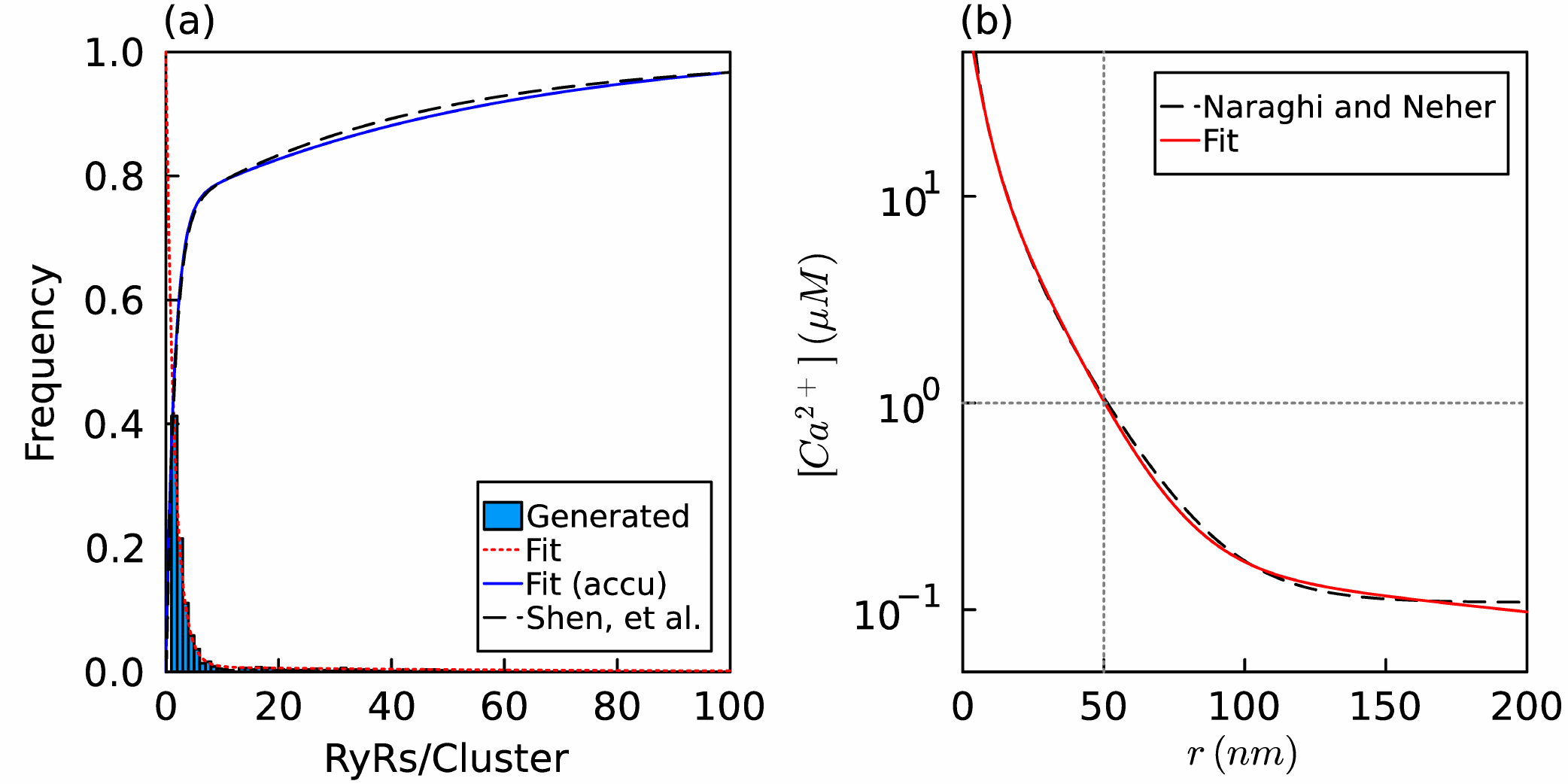}
\caption{(a) The frequency distribution of cluster size. The black dashed, blue full, and red dotted lines are for the experimental result by Shen $et~al.$~\cite{Shen2019}, fitted accumulated result $g(x)$, and fitted result $f(x)$. The histogram is for the randomly generated result.  (b) The calcium gradient against the distance. The black dashed and red full lines are for the simulation result in Ref.~\cite{Naraghi1997} and fitted result, respectively. }
\label{Fig: fit}
\end{figure}

The most important characteristic of spatial distribution of the calcium
channels is the cluster structure.  To construct a geometric network
reflecting the spatial distribution in a two-dimensional plane, the clusters should
be  randomly generated first. The observations in
Refs.~\cite{Baddeley2009,Jayasinghe2018,Shen2019} show that the clusters have
different sizes.  Hence, the  frequency distribution of cluster sizes is
required to generate a series of cluster sizes  for further generation of clusters. There exist several experimental results about the frequency
distribution of cluster sizes in the
literature~\cite{Baddeley2009,Jayasinghe2018,Shen2019}. Generally speaking, the
number of  smaller clusters is much larger than these with large size, and
the frequency distribution exhibits a power law, that is, the frequency decreases
exponentially with the increase of cluster size.  In simulations, we adopt the
recent experimental data for a two-dimensional surface by Shen
$et~al.$~\cite{Shen2019} as a black dashed line in
Fig.~\ref{Fig: fit}. To randomly generate the cluster sizes, we fit the
experimental data with an integral of exponential functions as, 
\begin{equation}
g(x)=\int^x_0f(y)dy=\int^x_0(ae^{-by}+ce^{-dy})dy.
\end{equation}
The experimental  data of the accumulated frequency distribution are well-fitted,
and shown as the  blue solid line in Fig.~\ref{Fig: fit}. The normalized function of
frequency distribution can be obtained as,
\begin{equation}
f(x)=0.991e^{-0.66 x} + 0.009 e^{-0.017 x}.   
\end{equation}
The first term corresponds to the power law of the frequency distribution. The
second term is for an enhancement of small size clusters, which has been
observed in many experiments~\cite{Baddeley2009,Shen2019}.

With the obtained frequency distribution, a series of cluster sizes can be
generated. In the current paper, we consider $N=4000$ channels. Two random
numbers are introduced as $r_1$ in [0,1] and $r_2$ in [0,$n_{\rm max}$]. Here, we
only consider the cluster sizes which are not too large to avoid the
uncertainty introduced by extremely large cluster in a limited area
considered. A maximum size of the cluster as $n_{\rm max}=100$ is adopted in
simulation.  It is also consistent with the experimental observation, in which
the clusters composed of more than 100 RyR's were scarcely
observed~\cite{Baddeley2009,Jayasinghe2018,Shen2019}.  If $r_1<f(r_2)$, the
$r_2$ will be recorded as a value of cluster size. If not, these two random
numbers are discarded.  Then, another two random numbers are generated and
judgement is repeated until the total number of RyR's reaches the total number of
channels $N$. The frequency distribution of generated 783 cluster sizes is also
presented and compared with the fitted function $f(x)$  in Fig.~\ref{Fig: fit}
(we would like to note that all clusters with size 0 are discarded). One can
find that the experimental frequency distribution is well generated with the
above method. 

\begin{figure*}[htbp!]
  \includegraphics[bb=0 0 900 500,scale=0.58 ,clip]{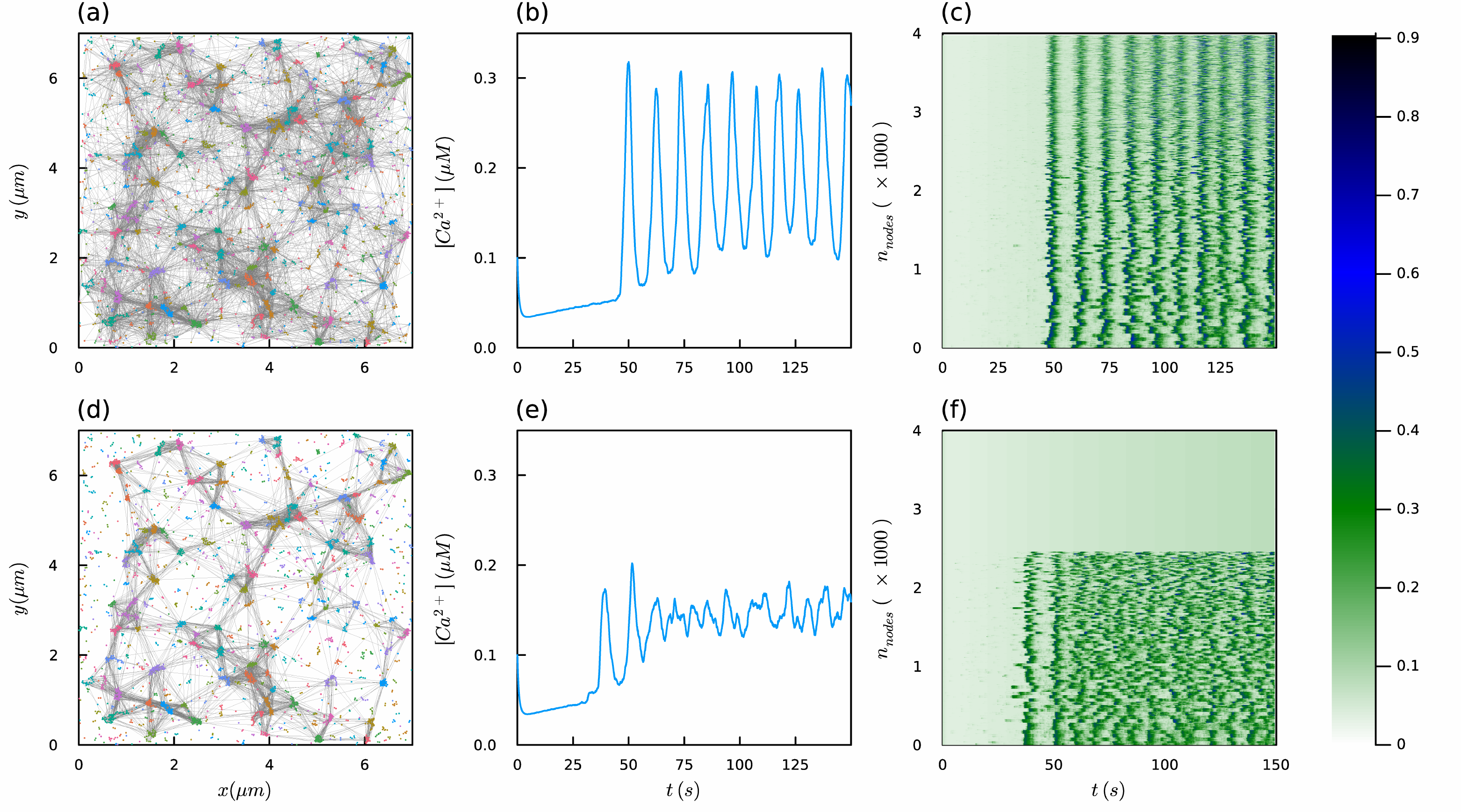}
\caption{Calcium oscillation on the geometric network. (a), (d) Spatial distribution of RyR's with random connections. (b), (e) Evolution of the calcium concentration $[{\rm Ca}^{2+}]$ with variation of time $t$. (c), (f) Evolution of the calcium concentration $[{\rm Ca}^{2+}]$ near each RyR with variation of time $t$. (a)-(c) The network with all clusters considered. (d)-(f) The network without clusters smaller than 10 RyR's. }
\label{Fig: 2D}
\end{figure*}

In the above step, a series of  cluster sizes are randomly generated. In the
followings,  clusters with these sizes will be generated.  In the literature,
the structure of a cluster is found to be complex.  In this paper, we adopt the picture
suggested by the super-resolution imaging in Ref.~\cite{Jayasinghe2018}. The
observations suggest that the cluster is not compact, and the shape is
irregular. Fortunately, in that paper, the authors also provided a method to
randomly generate a cluster, which is  adopted directly in the current paper.
The simulation starts with an original position ${\bm x}_1=(0,0)$, which is
recorded in a vector as the position of the first RyR.  Then, a random direction
$\theta$ is generated with a distance $r$ which is varied slightly around a mean
distance of 40 nm according to a Gaussian distribution with a sigma of 7.4 nm.
It closely matches the observed distance distribution in mean and width in
Ref.~\cite{Jayasinghe2018}. The position moves to ${\bm x}_2={\bm x}_2+\Delta
{\bm x}$ with a translation $\Delta {\bm x}=(r\cos\theta, r\sin\theta)$. The
position ${\bm x}_2$ is recorded in the vector as position of the second RyR.
Then, another random direction and distance are generated.  The new translation
is added into previous position ${\bm x}_2$,  and a new position ${\bm x}_3$ is
obtained and recorded again. The steps continue until the RyR position series of
a cluster with certain size is obtained and recorded in the vector. As shown in
Fig.~\ref{Fig: 2D}(a), this self-assembly process leads to the appearance of
some larger gaps in the clusters similar to those observed in
Ref.~\cite{Jayasinghe2018}. 

The randomly generated clusters with different sizes need to be scattered
randomly in a two-dimensional plane. In the literature~\cite{Chena2018}, the role
of the Rogue and clustered RyR's in a two-dimensional plane was also discussed, and
the rogue RyR's were found important in the generation of calcium sparks. In that
work, the clustered and Rogue RyR's were grouped into a CRU (Ca$^{2+}$ release
units).  However, in many
observations~\cite{Baddeley2009,Jayasinghe2018,Shen2019}, the clusters with
different sizes distribute randomly, which is adopted in the current paper.
A square region with a length of $l=d\sqrt{N_{\rm cluster}}$ is considered with
$d$ and $N_{\rm cluster}$ being the mean distance between two neighbor clusters and
total number of clusters. The distribution of nearest neighbor distances was
studied in the literature, and the largest possibility appears at about 200 nm with
a long tail~\cite{Jayasinghe2018}. In the current paper, we adopt a mean distance
as 250 nm.   For a cluster, two random numbers in range $[0, l]$ are generated
as the position of the cluster, which is added into the positions of all RyR's in
the cluster obtained in the previous step. If the clusters are spread randomly
in the two-dimensional plane directly, some clusters may overlap to each other. To
avoid such overlapping, we divide the two-dimensional square region into lattices
of 30$\times$30~nm$^2$. When the positions of the RyR's in a clusterare
determined, the occupied lattices are marked and can not be occupied again.
With such treatment, a random spread of the clusters is reached, and shown in
Fig.~\ref{Fig: 2D}(a).     

In our theoretical frame, the connections should be described by an adjacency
matrix. Hence, the last step to construct the geometric network is to abstract
the RyR distribution in a two-dimensional plane into nodes and edges of a network.
Obviously, the RyR's should be taken as nodes of the geometric network. In
Section~\ref{Network}, the edges are introduced according to the classical
networks where only the characteristics of the RyR distribution are considered.
In the current case, an explicit transition from the distance between a RyR pair
to a edge of network is required. The connection should reflect the magnitude of
variation of  calcium concentration. As discussed in Section~\ref{Mechanism},
the buildup of  calcium gradient between two RyR's is much faster than the
activation of RyR's~\cite{Soeller1997,Valent2007,Iaparov2021}.  Hence, the
connection of two RyR's can be determined by the calcium gradient arising from
the open channel.  The estimation of the calcium gradient is beyond the scope of
this paper. We adopt the simulation results in Ref.~\cite{Naraghi1997} as shown
in Fig.~\ref{Fig: fit}. It suggests a rapid decrease of the calcium
concentration with the distance between two channels, which becomes slower at
distance larger than about 100~nm. To make the further transition, it is fitted
in a range from 2~nm to 200~nm by three exponential functions as
\begin{equation} c(r)=77.7e^{-r/4.3}+21.7e^{-r/15.5}+0.19e^{-r/300},
\end{equation} 
where the $r$ is the distance  between two channels.
Here, we neglect very small distances where the calcium
concentration is very large. In fact, we connect all channel pairs with a
distance smaller than 50~nm, corresponding to 1~$\mu$M, which is smaller than
the size of two channels. For other channel pairs, the probability to have a
connection is determined by the calcium concentration with the following method.
A random number is generated, and compared with $c(r)$ with $r$ being the
distance of two channels. If the random number is smaller, the two channels will
be connected. The results are also presented in Fig.~\ref{Fig: 2D}(a).

\subsection{Simulation of oscillations on geometric network}

With the method in the above, a spatial distribution of 4000 RyR's in a
two-dimensional square region of 7$\times$7~$\mu$m$^2$ is generated as shown in
Fig.~\ref{Fig: 2D}(a). The RyR's are grouped into 783 clusters with different
sizes, which are non-compact, and have irregular shapes as in experimental observations.
A large amount of small clusters and rogue RyR's scatter around the large
clusters. The connections of the RyR's in the large clusters are dense. However,
different from the clique graph, the RyR's in a large cluster are not fully
connected  due to the irregular shape.  

The network in Fig.~\ref{Fig: 2D}(a) can be described with an adjacency matrix $A$
in Eq.~(\ref{Eq: fluxRyR}). The average degree of the obtained network is about
10.  Considering the total number of RyR's of 4000, the geometric network is a
sparse network. Such a network has a sparsity close to those of a Erd\"os-R\'enyi
network $G_{\rm ER}(1000, 0.002)$ in Fig.~\ref{Fig: ER}(g) and a BaraB\'asi-Albert
network $G_{\rm ER}(1000, 2)$ in Fig.~\ref{Fig: BA}(g), which can not produce a calcium
oscillation. With the adjacency matrix, the evolution of the calcium
concentration can be simulated with the same procedure as for the classical
networks. With the same parameters of the calcium transduction mechanism, one
can find that the calcium oscillation can be reproduced as shown in
Fig.~\ref{Fig: 2D}(b).  The evolution of calcium concentrations near 4000 RyR's
are presented in Fig.~\ref{Fig: 2D}(c). The results suggest that the
synchronization can be kept well. These results suggest that the geometric network
based on the experimental information has better performance than the
Erd\"os-R\'enyi or BaraB\'asi-Albert network with the same sparsity.

To understand the role of small clusters and rogue RyR's, the connections to RyR's
in all clusters with sizes smaller than 10 RyR's are removed as shown in
Fig.~\ref{Fig: 2D}(d).  For comparison with the results with all clusters, the
small clusters and rogue RyR's are kept in the figure. One can find that
the sketch of the connections between clusters is still kept.  Although the large
clusters are still connected, as shown in Fig.~\ref{Fig: 2D}(e), the calcium
oscillation disappears after removing  the small clusters. In Fig.~\ref{Fig:
2D}(f), the time revolution of every RyR is illustrated. No oscillation happens
near the rogue RyR's and small clusters whose connections are removed. The
oscillation  still can be seen for a single large cluster, though not so well.
However, the global calcium oscillation fails, as on the clique graph network, due
to the desynchronization of different clusters. The synchronization can not be
maintained only with connections between the large clusters. The results support
the importance of the small clusters and rogue RyR's in the formation and
maintenance of calcium oscillation.

\section{Summary and discussion}\label{Summary}

The calcium oscillation is a  physiological phenomenon, which is very important
to regulate intracellular life activity, and abnormal amplitude and frequency
have a close relationship to many diseases. The calcium channel is a key ingredient
in the calcium regulation mechanism to reproduce the calcium oscillation. Though
there exist many models which are successful to reproduce the oscillations, the
effect of the spatial distribution of calcium channels beyond the cluster is
scarcely investigated in existing studies. In this paper, we establish a
theoretical framework beyond the existing models to study the effect of the
spatial distribution on the calcium oscillation. The Keizer-Levine model is
introduced to describe the transition of four states of the calcium channel, and
calcium exchange between the cytoplasm and the stores or external medium.
Since the Keizer-Levine model is enough to successfully generate the calcium
oscillation, in the current paper the mechanism is extended directly from the
Keizer-Levine model under the mean-field ansatz to a network model.  The variation
of the calcium concentration induced by the open channels is described by an
adjacency matrix of considered networks with channels as nodes and Ca$^{2+}$
connections as edges. The theoretical frame is checked first by the complete
graph, and the mean-field result is well reproduced. 

Since the real spatial distribution of the calcium channels is very complex,
several types of classical networks, which reflect different sides of
characteristics of channel distribution, are introduced to perform simulation.
The Erd\"os-R\'enyi network reflects the homogeneous distribution at large
scale.  The heterogeneous BaraB\'asi-Albert network reflects  a power law of
cluster sizes, in which clusters are taken as nodes. For the Erd\"os-R\'enyi
network, if parameter $p$ is chosen larger than 0.1, the interspike interval and
amplitude are almost the same as the results in the mean-field approximation 
with some uncertainties from the randomness introduced in the simulation.
Different from the Erd\"os-R\'enyi network, the BaraB\'asi-Albert network
deviates from the mean-field approximation in almost a full range of the
parameters. If the parameters are justified, the BaraB\'asi-Albert network can
produce an oscillation with a similar frequency and amplitude as the
Erd\"os-R\'enyi network.  It may be the reason why the models under mean-field
ansatz are also successful to describe many experimental observations. However,
the Erd\"os-R\'enyi network lacks the regulation ability of the
BaraB\'asi-Albert network by changing the architectures. The modulation of
frequency and the amplitude of  calcium oscillation is important to intracellular
life activity~\cite{Pratt2020}, such as  sensitive and specific response of the
effector proteins~\cite{Dolmetsch1998}. For both the Erd\"os-R\'enyi and
BaraB\'asi-Albert networks which are not very sparse, the synchronization of the
nodes is very well, which makes global oscillation possible. If the network
becomes sparse, which means that the couplings between nodes are far from a
global coupling, the  synchronization is broken down with the increase of the
sparsity, as suggested in Ref.~\cite{Baspinar2021}.  With the increase of
sparsity of network, the frequency increases and the amplitude decreases.  For a
very sparse network, the synchronization, as well as oscillation, will
disappear. 

In the clique graph and geometric network, the cluster structure is considered
explicitly.  Cluster structure is a well-known and important characteristic of
channel distribution. Early observations suggest that RyR's in a cluster are
packed compactly, forming a tight lattice~\cite{Franzini-Armstrong1999}.  Such
a picture was also adopted in many studies focusing on the local
dynamics~\cite{Rudiger2010,Qi2014,Cao2013}.  To reflect such a characteristic, a
simulation is performed with a clique graph, in which the channels in a cluster
are fully connected. The results suggest that local oscillation and local
synchronization can be  reproduced in the cluster if its size is large enough.
The global oscillation does not exist due to the desynchronization between
clusters. It suggests that a  regular compact picture of clusters with a relatively
large distance between the clusters is not suitable to describe the realistic
channel distribution. Besides the clique graph, a geometric network is
constructed in a two-dimensional plane with the experimental information in
Refs.~\cite{Scott2021,Baddeley2009,Jayasinghe2018}.  The cluster sizes are
randomly generated with the experimental frequency distribution. The RyR
arrangements of a series of clusters with different sizes are simulated randomly
with the experimental mean distance. The calcium oscillation can be well
reproduced from the geometric network with the cluster structure.  As shown in
Fig.~\ref{Fig: 2D}(f), about two-thirds of the RyR's distribute in  clusters
larger than 10 RyR's. However, if the RyR's in small clusters are removed, the
synchronization between the clusters can not be maintained, though the sketch of
the connections is still kept. Such results suggest the importance of  small
clusters in the formation and maintenance of calcium oscillation.  

In summary, a theoretical framework is first established to study the spatial
distribution of calcium channels with complex structures by simulating the
calcium oscillation on three classical networks and a more realistic geometric
network. The mean-field result can be well reproduced in the simulations on the
homogeneous networks.  The power law of cluster sizes is referred to due to more
effectivity to modulate calcium oscillation.  The simple cluster model is
disfavored due to the desynchronization between the clusters. However, with the
small clusters included, the calcium oscillation can be reproduced.  The
recent super-resolution nanoscale imaging supports such
conclusion~\cite{Scott2021,Baddeley2009,Jayasinghe2018}. 

The current research provides a helpful basis to construct a more realistic model to
study the channel distribution,  and can be extended to study other phenomena in
calcium signal transduction. However, in the current paper, the diffusion of
calcium is only included as the connection between calcium channels while the
delay time is neglected.  The diffusion of the free calcium should be 
compared with the open time of the RyR. Hence, the time delay between the RyR's can
safely be neglected for free calcium. The buffers, such as calmodulin, will
make calcium diffuse slower than  free calcium. Considering that the buffers
will carry calcium far from the endoplasmic reticulum, in the current model we
assume that the calcium once attached to buffers does not affect the calcium
channels. In fact, its effect can be partly absorbed into the influx and efflux,
which exchange calcium with an external medium. In the current paper, the effect of
the diffusion is not fully included. The effectiveness of such treatment needs
further studies, such as simulation of the calcium spark. In addition, the current
paper is based on the Keizer-Levine model, in which some ingredients are not
included, such as the variation and diffusion of the calcium in endoplasmic
reticulum. Also, the mechanism about buffers is also not introduced explicitly. To
extend the current framework to study a more experimental phenomena, more suitable
mechanism should be included to establish the network model. 

\vskip 10pt

\noindent {\bf Acknowledgement} This project is supported by the National Natural Science
Foundation of China (Grant No. 11675228).

\end{document}